\documentclass[12pt]{article}

\usepackage{amssymb}
\usepackage{amscd}
\usepackage{graphics}
\usepackage{amsfonts}
\usepackage{amsmath}
\usepackage{latexsym}
\usepackage[all]{xy}
\usepackage{fancyhdr}
\usepackage{color}
\usepackage{empheq}

\usepackage{amsthm}

\usepackage{latexsym,graphics,color}
\def\bes{\begin{subequations}}
\def\ees{\end{subequations}}
\def\ba{\begin{align}}
\def\ea{\end{align}}

\def\th{\theta}

\def\w{\wedge}
\def\ot{\otimes}
\def\js{\frac{1}{4}}
\def\dg{\dagger}

\def\be{\begin{equation}}
\def\ee{\end{equation}}
\def\D{\mathcal D}
\def\vz{\boldsymbol{Z}}
\def\A{\mathcal A}

\def\K{\mathcal K}
\def\bl{\color{blue}}

\def\E{\mathcal E}
\def\A{\mathcal A}
\def\B{\mathcal B}
\def\cpn{\mathbb C\mathbb P^N}

\def\bc{\mathbb C}
\def\br{\mathbb R}
\def\bL{\boldsymbol L}

\def\J{\mathcal J}
\def\F{\mathcal F}
\def\k{\kappa}
\def\s1{\sigma^1}
\def\s2{\sigma^2}
\def\s3{\sigma^3}
\def\st{\sin{\theta}}
\def\ct{\cos{\theta}}

\def\dg{\dagger}
\def\noi{\noindent}
\def\L{\mathcal L}

\def\si{\sigma}

\def\bi{\boldsymbol\chi}
\def\vxi{\boldsymbol\xi}
\def\d{\partial}
\def\jp{\frac{1}{2}}
\def\ri{{\mathrm i}}

\def\sq{\sqrt{1-\vert\bi\vert^2}}
\def\vk{\boldsymbol K}

\def\en{\boldsymbol{e}_{N+1}}

\def\Ad{{\rm Ad}}
\def\ad{{\rm ad}}
\def\bw{\boldsympol{w}}

\definecolor{lila}{rgb}{1,0.2,0.9}
\definecolor{brown}{rgb}{0.5,0.3,0.3}
\definecolor{turquoise}{rgb}{0.2,0.9,0.7}
\definecolor{Orange}{rgb}{0.93,0.44,0}           
\definecolor{GrayBlue}{rgb}{0.35,0.4,0.62}       
\definecolor{SeafoamGreen}{rgb}{0.54,0.71,0.50}  

\definecolor{darkorange}{cmyk}{.20,.50,.80,0}
\definecolor{lightorange}{cmyk}{.07,.37,.65,0}
\definecolor{darkpeagreen}{cmyk}{.50,.30,.50,0}
\definecolor{lightpeagreen}{cmyk}{.22,.20,.40,0}

\newtheorem{thm}{Veta}[section]

\theoremstyle{definition}

\theoremstyle{definition}
\newtheorem{rem}[thm]{Remark}

\theoremstyle{proposition}

\theoremstyle{definition}

\def\G{{\cal G}}  
\def\vy{\boldsymbol Y}

\def\W{{\cal W}}%
\def\ri{{\mathrm{i}}}                   %
\def\1{{\mbox{\boldmath $1$}}}          %
\def\tr{\mathrm{tr\,}}                  %
\def\e{\epsilon}  %
\def\ka{\kappa}
\def\lm{\lambda}                        %
\def\bt{\beta}                          %
\def\jp{\frac{1}{2}}                    %
\def\om{\omega}                         %
\def\al{\alpha}        
\def\ga{\gamma}                         %

\def\bw{\boldsymbol{w}}%
\def\cp{\mathbb C\mathbb P}                     %

\def\bl{\color{blue}}      
      
\definecolor{spec}{rgb}{0.0, 0.26, 0.15}
\def\bx{\boldsymbol{x}}

\def\bp{\boldsymbol{p}}

\def\bxi{\boldsymbol{\xi}}

\def\bxi{\boldsymbol{\xi}}

\def\bpm{\begin{pmatrix}}
\def\epm{\end{pmatrix}}

\DeclareMathSymbol{\Rho}{\mathalpha}{operators}{"50}

\begin{document}

\begin{flushright}
{}~
  
\end{flushright}

\vspace{1cm}
\begin{center}
{\large \bf  Point particle $\E$-models}

\vspace{1cm}

{\small
{\bf Ctirad Klim\v{c}\'{\i}k}
\\
Aix Marseille Universit\'e, CNRS, Centrale Marseille\\ I2M, UMR 7373\\ 13453 Marseille, France}
\end{center}

\vspace{0.5 cm}

\centerline{\bf Abstract}
\vspace{0.5 cm}
\noindent  We show that the same algebraic data that permit to construct the Lax pair and the $r$-matrix of an integrable non-linear $\si$-model in $1+1$ dimensions  can be also used for the construction of Lax pairs and of $r$-matrices of several other non-trivial integrable theories in $1+0$ dimension. We call those new integrable theories the point particle $\E$-models, we describe their structure and  give their physical interpretation.   We work out in detail the point particle $\E$-models
associated to the bi-Yang-Baxter deformation of the $SU(N)$
principal chiral model. In particular, for  each complex  flag manifold we thus obtain a two-parameter family
of integrable models living on it.

  \vspace{2pc}
  \section{Introduction}

Integrable  dynamical systems with finite and with infinite number degrees of freedom  represent two related areas of research which  
often influenced each other in the past. 
The finite case is certainly  more tractable analytically, but  this fact does not really  make easier the task to construct new integrable models  comparing with the same task in the infinite case. Indeed, to identify the  Lax pairs and the   $r$-matrices of the models represent a tough problem in the both, finite and infinite, cases, and one may even argue that
it was the progress in the infinite
case in the sixties of the last century that changed the game also  in the finite case and made possible to go beyond a few examples with finite (and low)  number of degrees of freedom that had been known before. We stress at this point that the influence of the infinite on the finite was not merely at the level of a sort of inspiration but there  exist    explicit quantitative bridges linking the infinite to the finite. As an example of that situation we may mention the dynamics of a finite number of solitons of some infinite dimensional integrable field theories which
   turns  out to be captured by the dynamics of the integrable
   systems with  finite numbers of degrees of freedom, like the celebrated
    Calogero  or  Ruijsenaars-Schneider models \cite{C,RS}.
  
  \medskip
 
Nowadays, the integrability is still a very active domain of research, as we may wittness looking e.g. at  the spectacular recent  progress achieved in the field of integrable nonlinear $\si$-models describing string dynamics in curved backgrounds \cite{K08,Sf,DMV,K14,DLMV,BW,DHKM,DHT,DMV15,SST,DST}.
In the present article, we construct another bridge from the infinite to the finite, by associating  nontrivial integrable systems with   finite number  of degrees of freedom to   the recently constructed integrable    nonlinear $\sigma$-models in $1+1$ dimensions.

 \medskip 

 It turns out that the first order Hamiltonian dynamics of the most (if not all) of the known integrable nonlinear $\si$-models is expressible in terms of the so-called $\E$-models. Recall that the  $\E$-model  is a first order Hamiltonian dynamical system originally introduced in \cite{KS95} in the context   of generalized T-dualities in string theory and, somewhat surprisingly, it became clear only much later how useful is this concept also in the theory of integrable models
 \cite{K15,K17,K19,LV20,L21,LV23,H,HL}. As its name suggests, the $\E$-model
  is constructed out of a quantity denoted by $\E$, which is actually any involutive symmetric linear operator  acting  on the Lie algebra $\D$ of certain  quadratic Lie group $D$ referred to as the Drinfeld double. The symplectic form $\om$ and the Hamiltonian $H_\E$ of the   $\E$-model are given by the formulas\footnote{Strictly speaking, the phase space of the $\E$-model is the homogeneous space $LD/D$ and the  form $\om$ on $LD$ given by \eqref{symfo} is the pull-back of the true symplectic form by the projection map $LD\to LD/D$. }
  \be   \omega = -\jp(l^{-1}dl\stackrel{\wedge}{,}(l^{-1}d l)')_{L\D}\label{symfo},\ee
 \be H_\E  = \jp (l'l^{-1},\E \ \! l'l^{-1})_{L\D}.\label{Hamo}\ee
 Here  $l=l(\si)$ is an element of the loop group $LD$ and the prime denotes a particular derivative on the loop Lie algebra $L\D$ which is actually the standard derivative $\d_\si$ with respect to the loop parameter $\si$. Furthermore, $l^{-1}dl$ is the left-invariant Maurer-Cartan form on   $LD$, $(.,.)_{L\D}\equiv \oint d\si(.,.)_\D $ is the non-degenerate ad-invariant symmetric bilinear form on $L\D$   
where $(.,.)_\D$ is the canonical bilinear form with the same properties on $\D$.

\medskip 

Not every $\E$-model is integrable. However,  if the data $D,\E$ fulfil  
some sufficient conditions found in \cite{S,K21} then the 
integrability is guaranteed, that is, one can construct
the Lax  matrix and the $r$-matrix of the $\E$-model $(\om,H_\E)$.
Those sufficient conditions are quite simple and, what is essential for us, they have {\it finite} dimensional Lie algebraic character. Namely, it is required the existence of an one-parameter family of linear maps\footnote{If the spectral parameter $\lm$  takes a complex value,
 the   map $O(\lm)$ is considered as the map from
  $\D$ to $\G^\bc$.} $O(\lm):\D\to\G$ verifying
 \be [O(\lm)x, O(\lm)\E x]_\G=O(\lm)[x,\E x]_\D, \quad \forall x\in\D\label{sufco}\ee
 and of a two-parametric family of linear operators $\hat r(\lm,\rho):\G\to\G$ verifying
  \be [O^\dagger(\lm)x,O^\dagger(\rho)y]_\D+O^\dagger(\lm)[x,\hat r(\lm,\rho)y]_\G+O^\dagger(\rho)[\hat r(\rho,\lm)x,y]_\G=0,\quad \forall x,y\in\G\label{1do}\ee
  \be (O^\dagger(\lm)x,O^\dagger(\rho)y)_\D+(x,\hat r(\lm,\rho)y)_\G+(\hat r(\rho,\lm)x,y)_\G=0,\quad \forall x,y\in\G.\label{2do}\ee
   Here   $\G$ is a Lie subalgebra of $\D$ equipped with a quadratic structure $(.,.)_\G$ that is not necessarily induced by the quadratic structure $(.,.)_\D$, and $O(\lm)^\dagger:\G\to\D$
 is  the adjoint  of the operator $O(\lm)$  defined by the relation
 $$  (O(\lm)x,y)_\G=(x,O^\dagger(\lm)y)_\D, \quad  \forall x\in\D,\  y\in\G.$$

\medskip

By now many solutions of the sufficient conditions \eqref{sufco}, \eqref{1do} and  \eqref{2do} are known, but our goal here
is not to find yet some new ones. Instead, we want to show that to every known solution
$D,\E,O(\lm),\hat r(\lm,\rho)$
of those sufficient conditions we may associate not just one integrable $\E$-model living on the loop group homogeneous space $LD/D$ but also many other integrable 
theories with finite number of degrees of freedom living on appropriate homogeneous spaces of the group $D$. How this comes about?

\medskip

Let $D,\E,O(\lm),\hat r(\lm,\rho)$ be a solution of the sufficient conditions
\eqref{sufco},\eqref{1do} and \eqref{2do}
and pick an element $\xi\in\G\subset \D$ such that it holds
\be \ad_\xi^\G r(\lm,\rho)-r(\lm,\rho)\ad_\xi^\G =0,\quad \ad_\xi^\G O(\lm)-O(\lm)\ad^\D_\xi =0.\label{suc}
\ee
Then we can construct an integrable dynamical system living
on a  finite dimensional  phase space which has the structure of the  homogeneous space
$D/D_\xi$, where $D_\xi$ is the stabilizer of $\xi$ by the adjoint action of $D$.  
The easiest way to describe the symplectic form and the Hamiltonian of this finite dynamical system is in terms of their pullbacks $\omega^\xi$ and $H^\xi_\E$ with respect to the projection map $D\to D/D_\xi$. They read
\be   \omega^\xi =-\jp  \bigl(l^{-1}dl\stackrel{\wedge}{,}(l^{-1}dl)'\bigr)_{\D}\label{isymf},\ee
 \be H^\xi_\E  =\jp \bigl(l'l^{-1},\E \ \! l'l^{-1}\bigr)_{\D},\label{iHam}\ee
where
$$  l'l^{-1}:=\xi-\Ad_l\xi, \quad (l^{-1}dl)':=\ad_\xi(l^{-1}dl).$$ 
 By an abuse of notation, we shall refer to $\omega^\xi$ and $H^\xi_\E$ as the symplectic form and the Hamiltonian and
we shall call the dynamical system
$(D/D_\xi,\omega^\xi,H^\xi_\E)$ 
the {\it point particle $\E$-model} because of the similarity of the formulas
\eqref{isymf},\eqref{iHam} with \eqref{symfo},\eqref{Hamo}.

\medskip 

Given the integrable string data $D,\E,O(\lm),\hat r(\lm,\rho)$, there are typically  many solutions $\xi$ of the supplementary conditions
\eqref{suc}. The point particle $\E$-models based on the same data $D,\E,O(\lm),\hat r(\lm,\rho)$ but different choices of $\xi$ are often dynamically nonequivalent.  
  
\medskip 

The $\E$-model, whether the 
standard string one living on $LD/D$ or the point particle one living on $D/D_\xi$, is the first order Hamiltonian dynamical system.
In order to give it a physical interpretation, we have to postulate which coordinates of the phase space play the role of coordinates on the configuration space and which are generalized  momenta. In the stringy
$LD/D$ case, the momenta are parametrized via a choice of maximally isotropic subgroup $H$ of the Drinfeld double $D$ while the coordinates on the configuration space are expressed in terms of the functions on the space of cosets $D/H$.
Said differently, the stringy $\E$-model is interpreted as the non-linear $\si$-model on the target space $D/H$, where the geometry of this target
can be unambiguously  extracted from the data $(D,H,\E)$ (see \cite{KS,K95,KS96,K21} for details). 
 
\medskip

In the point particle $D/D_\xi$ case, the physical
interpretation may be also achieved by choosing a maximally isotropic subgroup $H$ of the Drinfeld double $D$, but the geometry of the configuration space, on which the point particle moves, depends on the choice of $\xi$. In Sections 5 and 6, we discuss several concrete examples working with   two particular Drinfeld doubles of the group $SU(N)$ and with various choices of $\xi\in{\rm Lie}(SU(N))$.  In all those cases the configuration space turns out to be   an appropriate complex flag manifold.  We do not provide  all details here in the introduction, but  we do give  a flavour of what is going on  by considering the Drinfeld double $D=T^*SU(2)$  and $\xi$ being the generator of the Cartan subalgebra of $su(2)$.
Thus, for a suitable choice of $\E$,  the configuration space turns out to be the standard sphere $S^2$ and the corresponding second order action of the point particle $\E$-model is
 \be S=\frac{1}{2}\int dt \left(\dot x_1^2+\dot x_2^2+\dot x_3^2 +4x_3 \right),\quad x_1^2+x_2^2+ x_3^2=1.\label{speo}\ee 
We recognize in \eqref{speo} the action of the spherical mathematical pendulum in the homogeneous
gravitational field which is of course known to be integrable. For comparison, in the string $LD/D$ case, the same Drinfeld double 
$D=T^*SU(2)$ and the same operator $\E$ turn out to give rise to the principal chiral model on the group $SU(2)$ governed by the second order action
\be  S=\jp\int d\tau d\si \tr\left(-(k^{-1}\dot k)^2+(k^{-1}k')^2\right)  =\int d\tau d\si \sum_{j=0}^3(\d_\tau n_j\d_\tau n_j-\d_\si n_j\d_\si n_j),\label{o4}\ee
$$ k= \bpm n_0+in_3 &-n_1+\ri n_2\cr n_1+\ri n_2&n_0-\ri n_3\epm, \quad n_0^2+n_1^2+n_2^2+ n_3^2=1.$$

 \medskip 

 Comparing the actions \eqref{speo} and \eqref{o4}, we 
 observe the the former is not   the standard  dimensional reduction of the latter. In particular, there is a potential term in \eqref{speo}, that the dimensional reduction cannot produce. 
This is consistent because what we do here is not  
the dimensional reduction of two-dimensional theories. Indeed,  we do not let the space derivative 
terms vanish but we rather replace them with the commutators with
$\xi$. In particular, this  yields   the potential term in \eqref{speo}. The "raison d'\^etre" of our procedure resides in the fact that we thus obtain the mechanical  Lax pairs 
from the field-theoretical ones automatically.

 \medskip

 The plan of the paper is as follows: In Section 2, we   remind the structure of the $LD/D$ string  $\E$-models and, in Section 3, we describe in detail the structure of the 
$D/D_\xi$ point particle $\E$-models. Then in Section 4,  we show how to construct the Lax operators and the $r$-matrices of  the integrable point particle $\E$-models  out of the solutions $ O(\lm),\hat r(\lm,\rho)$ of the sufficient conditions \eqref{sufco},\eqref{1do} and \eqref{2do}.  Then
we work out examples of the point particle $\E$-models based on the Drinfeld doubles 
of  the group $SU(N)$; in Section 5 we work with the double $T^*SU(N)$ while  in Section 6  with the Lu-Weinstein double $SL(N,\bc)$.
In the both cases, we work out  the second order actions on the configuration spaces which turn out to correspond to the motion of point particles on flag manifolds. We finish by conclusions and perspectives.

 \section{Reminder:  stringy $\E$-models}
   Recall that the Drinfeld double $D$ is a connected even-dimensional
 Lie group   equipped with a bi-invariant pseudo-Riemannian metric of maximally-Lorentzian (split) signature.
 This pseudo-Riemannian metric   naturally induces the non-degenerate ad-invariant symmetric bilinear form $(.,.)_\D$ defined on the Lie algebra $\D$ of $D$.

  A stringy $\E$-model, introduced in \cite{KS95,K15}, is  the first-order Hamiltonian dynamical system $(\omega,H_\E)$ living on the loop group  homogeneous space $LD/D$.    The  pull-backs of the symplectic form  and  of  the Hamiltonian from $LD/D$ on $LD$ are respectively given by the
 formulas
  \be   \omega = -\jp(l^{-1}dl\stackrel{\wedge}{,}(l^{-1}d l)')_{L\D}\label{symf},\ee
 \be H_\E  = \jp (l'l^{-1},\E \ \! l'l^{-1})_{L\D}.\label{Ham}\ee
 Here  $l=l(\si)$ is an element of the loop group $LD$ and the prime denotes the  derivative $\d_\si$ with respect to the loop parameter $\si$. Furthermore, $l^{-1}dl$ is the left-invariant Maurer-Cartan form on $LD$  and  $(.,.)_{L\D}\equiv \oint d\si(.,.)_\D $ is the non-degenerate ad-invariant symmetric bilinear form on $L\D$. Finally, $\E:\D\to\D$ is the  $\br$-linear operator squaring to identity, symmetric with respect to the bilinear form $(.,.)_\D$ and such that the bilinear form $(.,\E.)_\D$
 on $\D$ is strictly positive definite.

  \medskip
 
 It turns out, that the left action  of the group $LD$ on $LD/D$ 
 is generated by the moment map $j=l'l^{-1}$.
 The components of the current  $j$ then verify the Poisson current algebra
 \be \{(j(\sigma_1),T_1)_\D,(j(\sigma_2),T_2)_\D\}=(j(\sigma_1),[T_1,T_2])_\D\delta(\sigma_1-\sigma_2)+(T_1,T_2)_\D\delta'(\sigma_1-\sigma_2)\label{spc}\ee
 which plays a crucial role in the dynamics of the $\E$-models. Indeed, the Poisson brackets $\eqref{spc}$ as well as the explicit form \eqref{Ham} of the Hamiltonian give the following
 first order equations of motion of the $\E$-model
 $$ \frac{\d j}{\d \tau}=\{j,H_\E\}=(\E j)'+[\E j,j]. $$
  { \section{Point particle $\E$-models}
 
 To every stringy $\E$-model, we associate a family of dynamical systems living in $1+0$ dimension
 which we shall call  the point particle $\E$-models.
 Given the stringy $\E$-model and its underlying Drinfeld double $D$, each member of this associated family   corresponds to some choice of an element $\xi$ from  the Lie algebra $\D$ of the double. Thus, if $\xi\in\D$,
 the symplectic form and the Hamiltonian of the corresponding point particle $\E$-model are given by the formulas (see for comparison \eqref{symf} and \eqref{Ham})
    \be   \omega^\xi =-\jp  \bigl(l^{-1}dl\stackrel{\wedge}{,}(l^{-1}dl)'\bigr)_{\D}\label{xisymf},\ee
 \be H^\xi_\E  =\jp \bigl(l'l^{-1},\E \ \! l'l^{-1}\bigr)_{\D},\label{xiHam}\ee
 where  now
 $$ l'l^{-1}:=\xi-\Ad_l\xi, \quad (l^{-1}dl)':=\ad_\xi(l^{-1}dl).$$
   
 \medskip
 
 Whatever $\xi\in\D$ we choose, we observe that the Drinfeld-Kirillov form
 \eqref{xisymf} is closed because it can be written
 as
 \be \omega^\xi=-d(\xi, l^{-1}dl)_\D.\label{xipol}\ee
 However, the form $\omega^\xi$ is only presymplectic since it is never non-degenerate. Indeed, $\omega^\xi$ is the pull-back of a closed non-degenerate (hence symplectic) form $\om^\xi_c$ living on the space of cosets $D/D_{\xi}$, where
 the subgroup $D_\xi\subset D$ stabilizes $\xi$ under the adjoint action of $D$ on $\D$. 
 We have thus
 $$ \om^\xi=\pi_\xi^*\om^\xi_c,$$
 where $\pi_\xi:D\to D/D_\xi$ is  the projection mapping. 
 
 Note  also, that the Hamiltonian $H^\xi_\E$ is invariant
 with respect to the action of the stabilizer $D_\xi$, therefore it exists a function $H^\xi_{\E,c}  $ living on the space of cosets $D/D_{\xi}$ such that
  $$ H^\xi_{\E} =\pi_\xi^*H^\xi_{\E,c} .$$
 All in all, the
 $\xi$-version of the point particle $\E$-model is the finite-dimensional dynamical system living on the symplectic manifold
 $D/D_\xi$, having the symplectic form $\om^\xi_c$ and the Hamiltonian $H^\xi_{\E,c}$. However,
 somewhat abusively, we shall often continue to view $D$ as the phase space and
  the quantities $\om^\xi$ and $H^\xi_{\E}$ living on $D$ as the symplectic form and the Hamiltonian of the $\xi$-version of the point particle $\E$-model.
 
 \medskip
 
 It turns out, that the left action $\nabla^L$ of the group $D$ on the homogeneous space $D/D_\xi$
 is generated by the moment map $j=l'l^{-1}$.
 This means that it holds
 \be \nabla^L_Tf:=\left(\frac{d}{ds}\right)_{s=0}f(e^{sT}l)=\{f,(j,T)_\D\}_\xi,\quad T\in\D\label{mm}\ee
 where $f$ is a function on  $D/D_\xi$ and
 $\{.,.\}_\xi$ is a Poisson bracket induced by the symplectic form $\omega_c^\xi$. In particular, we obtain from the bracket \eqref{mm} the following Poisson current algebra (cf. the formula \eqref{spc})
 \be \{(j,T_1)_\D,(j,T_2)_\D\}_\xi=(j,[T_1,T_2])_\D+(T_1,[\xi,T_2])_\D. \label{xicur}\ee
 As in the string case, the first order equations of motion of the point particle $\E$-model is
  \be \frac{d j}{d \tau}=\{j,H_\E^\xi\}_\xi=(\E j)'+[\E j,j], \label{xife}\ee
 but now we have $(\E j)'=[\xi,\E j]$.}
\section{Integrable point particle $\E$-models}

  In this section, we deal with  integrable point particle   $\E$-models $(D/D_\xi, \om^\xi, H_\E^\xi)$. The most important conclusion of the present article is the fact 
  that  the stringy   sufficient conditions of integrability \eqref{sufco},\eqref{1do},\eqref{2do}  guarantee the integrability also
in the point particle case provided they are supplemented by two more conditions \eqref{sucb}, \eqref{suc2b} involving the element $\xi$.  We gather all those conditions as follows
 
 \medskip

 \noi 1) It exists a one-parameter family of linear maps $O(\lm):\D\to\G$ which verifies
   \be \quad \ad_\xi^\G O(\lm)-O(\lm)\ad^\D_\xi =0,\label{sucb}
\ee
 \be [O(\lm)x, O(\lm)\E x]_\G=O(\lm)[x,\E x]_\D, \quad \forall x\in\D\label{sufcb},\ee where   $\G$ is a  Lie subalgebra of $\D$ such that $\xi\in\G$; 

 \smallskip 
 
\noi 2) It exists   a two-parametric family of linear operators $\hat r(\lm,\rho):\G\to\G$ which verifies 
\be \ad_\xi^\G r(\lm,\rho)-r(\lm,\rho)\ad_\xi^\G =0,\label{suc2b}\ee
 \be [O^\dagger(\lm)x,O^\dagger(\rho)y]_\D+O^\dagger(\lm)[x,\hat r(\lm,\rho)y]_\G+O^\dagger(\rho)[\hat r(\rho,\lm)x,y]_\G=0,\quad \forall x,y\in\G\label{1db}\ee
  \be (O^\dagger(\lm)x,O^\dagger(\rho)y)_\D+(x,\hat r(\lm,\rho)y)_\G+(\hat r(\rho,\lm)x,y)_\G=0,\quad \forall x,y\in\G,\label{2db}\ee
where   the quadratic structure $(.,.)_\G$ on $\G$ is not necessarily induced by the quadratic structure $(.,.)_\D$ on $\D$ and $O(\lm)^\dagger:\G\to\D$ is  the adjoint  of the operator $O(\lm)$  defined by the relation
  $$ (O(\lm)x,y)_\G=(x,O^\dagger(\lm)y)_\D, \quad  \forall x\in\D, y\in\G.$$
  
\medskip  
 
 Having a solution $O(\lm),\hat r(\lm,\rho)$ of the sufficient
 conditions 1), 2),  we must show how to construct out of it the Lax pair $L(\lm),M(\lm)$ and the dynamical $r$-matrix $r(\lm,\rho)$ governing the Poisson brackets of the matrix elements of the Lax matrix $L(\lm)$. We do it similarly as in the string case \cite{S,K21} by postulating

  \be L(\lm)  = \xi- O(\lm)j,\label{Laxpp}\ee
    \be M(\lm)=-  O(\lm)\E j.\label{ppM}\ee
   Using \eqref{sufcb}, it can be checked easily, that the  equations of motion \eqref{xife} can be represented in the Lax form
 with   spectral parameter
        \be  \{L(\lm),H_\E\}_\xi= \frac{dL(\lm)}{dt} =[L(\lm),M(\lm)]_\G.\label{ppLax}\ee
        
        Now set
           \be r(\xi,\rho)=C_{AB}(\hat r(\lm,\rho)T^A)\otimes {T^B},\label{ans}\ee
  where $C_{AB}$ is the inverse matrix of the matrix $C^{AB}$ defined by 
     \be C^{AB}:=(T^A,T^B)_\G,\ee
and $T^A$ is some basis of the Lie algebra $\G$ on the choice of which actually the $r$-matrix \eqref{ans} does not depend.

\medskip

We wish to show, that the quantity $r(\xi,\rho)$ defined by \eqref{ans} is the
  $r$-matrix of the point particle $\E$-model, which, following the general theory of the Lax
  integrable systems (cf.  \cite{BBT}), means the validity of the following fundamental relation
      \be \{L(\lm)\stackrel{\otimes}{,} L(\rho)\}_\xi=[r(\lm,\rho),L(\lm)\otimes {\rm Id}]-[r^p(\rho,\lm),{\rm Id}\otimes L(\rho)].\label{xisi}\ee
Here the notation $r^p$ means  
$$ r^p=\sum_\alpha B_\alpha\otimes A_\alpha,$$
provided  $r$ has the form
  $$ r=\sum_\alpha A_\alpha\otimes B_\alpha $$
  for some family of elements $A_\alpha,B_\alpha\in\G$.  
  
  \medskip

  \medskip
  
To see that the relation \eqref{xisi} is indeed satisfied, we calculate its matrix components.
On the left-hand-side, we obtain
 $$\{(x,L(\lm))_\G,(y,L(\rho))_\G\}=\{(x,O(\lm)j)_\G,(y,O(\rho)j)_\G\}=$$\be =(j,[O^\dagger(\lm)x,O^\dagger(\rho)y]_\D)_\D  +(O^\dagger(\lm)x,[\xi,O^\dagger(\rho)y]_\D)_\D, \label{curfour} \quad x,y\in\G, \ee 
 where we have used the Poisson brackets \eqref{xicur}. 
 On  the right-hand-side, we obtain   
\be  -(O(\lm)j,[x,\hat r(\lm,\rho)y])_\G-(O(\rho)j,[\hat r(\rho,\lm)x,y])_\G  +([\xi,x],\hat r(\lm,\rho)y)_\G-([\xi,y],\hat r(\rho,\lm)x)_\G.\label{xipor}\ee
  Now because of the conditions \eqref{1db},\eqref{2db} and \eqref{sucb}, the expression \eqref{xipor} matches precisely  
 \eqref{curfour}.

 \medskip

  The validity of the Lax equation
  \eqref{ppLax} and of the relation
  \eqref{xisi} means by definition the integrability of the model.

  \smallskip

 {
 \section{The Drinfeld double $T^*K$}
 \subsection{Point particle principal chiral model}

Let $K$ be a simple compact connected and simply connected Lie group and denote by $(.,.)_\K$ the standard Killing-Cartan form on its Lie algebra $\K$. Then the  cotangent bundle $T^*K$ of the Lie group  $K$ can be given the structure of the Drinfeld double $D$.  To see this, parametrize the elements of the cotangent bundle 
 by the pairs $(k,\ka)$, $k\in K$, $\ka\in \K$. If $1$ stands for the unit element of $K$ and $0$ is the neutral element of $\K$ (it is actually the unit element of the additive Abelian group underlying the vector space $\K$) then  the unit element of $D=T^*K$ is 
      $$ e=(1,0).$$
 Furthermore, the inverse element is
     $$ (k,\k)^{-1}=(k^{-1},-\Ad_{k^{-1}}\k).$$
      and the group  multiplication law reads
     $$ (k_1,\ka_1)(k_2,\ka_2)=(k_1k_2,\kappa_1+{\rm Ad}_{k_1}\kappa_2), \qquad k_1,k_2\in K,\quad \kappa_1,\kappa_2\in\K.$$
The symmetric non-degenerate ad-invariant  bilinear form $(.,.)_\D$ on the Lie algebra $\D$ is given by
   $$ \Bigl((\mu_1,\nu_1),(\mu_2,\nu_2)\Bigr)_\D=(\mu_1,\nu_2)_\K+
   (\mu_2,\nu_1)_\K,\quad \mu_{1,2},\nu_{1,2}\in\K,$$
   where $(.,.)_\K$ is the standard Killing-Cartan form on $\K$.
   The form $(.,.)_\K$ is strictly negative definite while the form $(.,.)_\D$ has the split signature 
   $(+,\dots,+,-,\dots,-)$.   The ad-invariance can be checked by using the following formula expressing the adjoint action of $D$ on $\D$
   \be \Ad_{(k,\k)}(\mu,\nu)=(\Ad_k \mu,\Ad_k \nu+\ad_\k\Ad_k\mu).\label{ada}\ee}
    
 \medskip

{ We may now define a natural $\E$-model on the double $D=T^*K$ by defining the operator  $\E:\D\to\D$  as 
   \be \E(\mu,\nu)=  -( \nu, \mu) .\label{eeta}\ee
   It is evident that $\E^2=$Id, moreover
   the form $(.,\E .)_\D$ is positive definite since
    \be \Bigl((\mu_1,\nu_1),\E(\mu_2,\nu_2)\Bigr)_\D= -(\mu_1,\mu_2)_\K -(\nu_1,\nu_2)_\K.\label{pos}\ee
   Finally, the formula \eqref{pos} implies also the symmetry of $\E$
    $$ \Bigl(\E(\mu_1,\nu_1),(\mu_2,\nu_2)\Bigr)_\D= \Bigl((\mu_1,\nu_1),\E(\mu_2,\nu_2)\Bigr)_\D.$$}
    
{ Let $\zeta\in\K$ be an element of the Lie algebra $\K$ and denote by $K_\zeta$ its stabilizer  under the adjoint action of $K$ on $\K$. It is then easy to derive from the formula \eqref{ada}, that the stabilizer $D_\xi$ of the element $\xi=(\zeta,0)$ is the
   cotangent bundle of the stabilizer $K_\zeta$
 $$ D_\xi=\{(k,\ka)\in T^*K, \quad k\in K_\zeta,
 \quad \ka\in\K_\zeta\}.$$
 In order calculate the Hamiltonian and the symplectic form
 of the point particle $\E$-model for  $D=T^*K$,
 $\E$ given by Eq.\eqref{eeta} and $\xi=(\zeta,0)$, it is convenient to parametrize the elements $l=T^*K$ as
 $$ l=(k,0)(1,\rho), \qquad k\in K,\quad \rho\in\K. $$
 Then we have
 \be j\equiv l'l^{-1}=\xi-\Ad_l\xi=(k'k^{-1},\Ad_k\rho'),\label{dec}   \ee
 where
 $$ k'k^{-1}:=\zeta-\Ad_k\zeta,\quad \rho':=\ad_\zeta\rho.$$
 Using the formula \eqref{xiHam}, we find
  \be H^\xi_\E=-\frac{1}{2}(k'k^{-1},k'k^{-1})_\K-\frac{1}{2\ }(\rho',\rho')_\K.\label{Hampc}\ee
 In order to calculate the
 symplectic form \eqref{xipol}, we need the formula for the left  
 Maurer-Cartan form on $D$
   $$l^{-1}dl\equiv(k,\k)^{-1}d(k,\k)=(k^{-1}dk,\Ad_{k^{-1}}(d\k)),$$
We obtain
  \be \omega^\xi=d(\rho',k^{-1}dk)_\K.\label{symfpc}\ee
  The formulas \eqref{Hampc} and \eqref{symfpc}  permit to write down the first order action principle for the point particle principal chiral model as
  \be S(k,\rho)=\int dt \Bigl( (\rho',k^{-1}\dot k)_\K+ \frac{1}{2}(k'k^{-1},k'k^{-1})_\K+\frac{1}{2 }(\rho',\rho')_\K\Bigr).\label{fopc}\ee
  Note that the variable $\rho'$ takes values in $\K^\perp_\zeta$. Varying the action \eqref{fopc} with respect to $\rho'$ gives
  \be \rho'=-P^\perp(k^{-1}\dot k),\label{ske}\ee
   where $P^\perp$ is  the orthogonal projector  which projects on the vector space $\K^\perp_\zeta$.
Inserting \eqref{ske} back into
  the action \eqref{fopc}  gives rise to the second order action
    \be S(k)=\frac{1}{2}\int dt \Bigl(- (P^\perp k^{-1}\dot k, P^\perp k^{-1}\dot k)_\K+  (k^{-1}k',k^{-1}k')_\K\Bigr),\label{sopc}\ee
    where 
    $$ k^{-1}k':=k^{-1}(k'k^{-1})k=k^{-1}\zeta k-\zeta.$$
Note that the action \eqref{sopc} describe the motion of a point particle on the space
    of cosets $K/K_\zeta$. Indeed, the action \eqref{sopc} has a gauge symmetry $k(t)\to k(t)h(t)$, where $h(t)$ is an arbitrary function of the time $t$ taking values in the stabilizer group $K_\zeta$.
    
    \medskip 

For comparison, the stringy $\E$-model based on the same data $D=T^*K$ and $\E$ given by \eqref{eeta} leads to the following second order action
 $$ S(k)=\frac{1}{2}\int dt d\si \Bigl(- ( k^{-1}\dot k, k^{-1}\dot k)_\K+  (k^{-1}k',k^{-1}k')_\K\Bigr).$$

 In particular, for $K=SU(2)$ and
 $\zeta=\bpm \ri&0\\0&-\ri\epm$ we obtain the actions listed already in the Introduction. In the point particle case it is that of the spherical pendulum in the homogeneous gravitational field 
 $$ S=\frac{1}{2}\int dt \left(\dot x_1^2+\dot x_2^2+\dot x_3^2 +4x_3 \right),\quad x_1^2+x_2^2+ x_3^2=1$$
and in the stringy case it is that of the $O(4)$ nonlinear $\si$-model in $1+1$-dimensions
$$ S= \int d\tau d\si \sum_{j=0}^3(\d_\tau n_j\d_\tau n_j-\d_\si n_j\d_\si n_j), 
 \quad n_0^2+n_1^2+n_2^2+ n_3^2=1.$$}
 
 { 
\subsection{Integrability}
   Let us show that the point particle principal chiral model is integrable. For that we  take  the Lie algebra $\G$ to be just $\K$ equipped with its standard Killing-Cartan
   form $(.,.)_\G=(.,.)_\K$. By the very logic of this article, we borrow  the required operators $O(\lm), \hat r(\lm,\rho)$ from the stringy principal chiral model. They  are given by \cite{DMV, K20}
$$ \hat r(\lm,\rho)=\frac{\rho^2}{1-\rho^2}\frac{1}{\rho-\lm}{\rm Id},\quad O(\lm)(\mu,\nu):=\frac{\mu-\lm\nu}{1-\lm^2}, \quad O^\dagger(\lm)\mu=  \left(\frac{-\mu\lm}{1-\lm^2}, \frac{\mu}{1-\lm^2}\right) $$
    and they therefore verify the conditions \eqref{sufcb},  \eqref{1db}, \eqref{2db}  automatically. We then check easily that the remaining conditions \eqref{sucb}, \eqref{suc2b} are also satisfied, so we conclude that the point particle principal chiral model \eqref{sopc} is integrable. 
    Using   \eqref{dec}, we find the corresponding 
    Lax pair \eqref{Laxpp} and \eqref{ppM}  
    \be L(\lm)= \zeta-\frac{1}{1-\lm^2} (k'k^{-1}-\lm k\rho'k^{-1}) ,\quad M(\lm)=-\frac{1}{1-\lm^2} (\lm 
 k'k^{-1}-  k\rho'k^{-1}).\label{laxp}\ee

j Taking into account the equation   \eqref{ske}, we can   rewrite the Lax pair \eqref{laxp} as
     $$ L(\lm)= \zeta- \frac{k'k^{-1}+\lm kP^\perp(k^{-1}\dot k)k^{-1}}{1-\lm^2} , \quad M(\lm)= - \frac{\lm k'k^{-1} +kP^\perp(k^{-1}\dot k)k^{-1}}{1-\lm^2}   $$
  and the Lax equation $\dot L=[L,M]_\K$ becomes
$$\frac{1}{1-\lm^2}\left(-\d_\tau{(k'k^{-1})}+\left( kP^\perp(k^{-1}\dot k)k^{-1}\right)'-[k'k^{-1},kP^\perp(k^{-1}\dot k)k^{-1}]\right)+$$\be +\frac{\lm}{1-\lm^2}\left(\left(k'k^{-1}\right)' -\d_\tau\left(kP^\perp(k^{-1}\dot k)k^{-1}\right)\right)=0.\label{ebo} \ee
Since \eqref{ebo} must hold for all values of $\lm$, we obtain two conditions to fulfill
\be -\d_\tau{(k'k^{-1})}+\left( kP^\perp(k^{-1}\dot k)k^{-1}\right)'-[k'k^{-1},kP^\perp(k^{-1}\dot k)k^{-1}]=0,\label{bia}\ee 
\be\left(k'k^{-1}\right)' -\d_\tau\left(kP^\perp(k^{-1}\dot k)k^{-1}\right)=0.\label{te}\ee
In fact, the first condition \eqref{bia} is fulfilled automatically, it is the so called Bianchi identity.
The second condition \eqref{te} is the  second order equation of motion   of the point particle principal chiral model.

\medskip

\subsection{Point particle on  $\cpn$}    
  It is instructive to work out  the point particle principal chiral model for  the case of the group $K=SU(N+1)$ and of the element
    \be \zeta=\ri \bpm \1_N &0\\0&-N\epm,\label{cpz}\ee where $\1_N$ stands for the unit $N\times N$ matrix. In this case the stabilizer group $K_\zeta$ is the biggest possible, it  has the structure
    $S(U(N)\times U(1))$, where the factor $U(1)$ is generated by the element $\zeta$ itself and $U(N)$ is formed by the   unitary matrices of the form
    $\bpm h&0\\0&1\epm$. The configuration space $K/K_\zeta$ of the models is thus the space of cosets
    $SU(N+1)/S(U(N)\times U(1))$ which is nothing but the complex projective space $\cpn$.
    
    \medskip 
    
    We now evaluate the second order action \eqref{sopc} on the $\cpn$ on the chart where the $(N+1)$-th  homogeneous coordinate $Z^{N+1}$ does not vanish. We can parametrize this chart by a complex $N$-vector $\bi$ of the norm strictly inferior to $1$
    so that the  homogeneous coordinates $Z^j$, $j=1,...,N+1$ 
    are of the form
    \be Z^1=\lm\chi^1,\quad Z^1=\lm\chi^2,\quad \dots , \quad Z^N=\lm\chi^N,\quad Z^{N+1}=\lm\sqrt{1-\vert\bi\vert^2},\label{hm} \ee 
    where $\lm$ is an arbitrary non-vanishing complex number. In what follows we  choose $\lm=1$, which implies, in particular
$$ \sum_{j=1}^{N+1}\bar Z^jZ^j=1.$$
    
    \medskip 
     
    We fix the gauge in \eqref{sopc} by setting $k=  \1_{N+1}$ for $\bi=\boldsymbol{0}$ and
  \be k=\bpm \1_N-\al \bi\otimes \bi^\dg & \bi\\ -\bi^\dagger&\sqrt{1-\vert\bi\vert^2}\epm,\quad \alpha=\frac{1-\sqrt{1-\vert\bi\vert^2}}{\vert \bi\vert^2},\quad \bi\neq\boldsymbol{0},\label{kchi}\ee
    which gives, in particular
 \be k^{-1}k'=(N+1)\ri\bpm -\bi\otimes\bi^\dagger &\sq\bi\\\sq\bi^\dagger&\vert\bi\vert^2,\epm
     \ee
    $$ P^\perp k^{-1}\dot k=
    \bpm \boldsymbol 0&\dot\bi-\dot{\sq}\bi-\al(\bi^\dagger\dot\bi)\bi\\-\dot\bi^\dagger+\dot{\sq}\bi^\dagger+\al(\dot\bi^\dagger\bi)\bi^\dagger&0\epm $$
 $$(k^{-1}k',k^{-1}k')_\K=-2(N+1)^2\vert\bi\vert^2.$$
    The point particle principal chiral model action \eqref{sopc}  therefore becomes
  \be S(\bi)=\int dt\left(\left\vert \dot\bi-\dot{\sq}\bi-\al(\bi^\dagger\dot\bi)\bi\right\vert^2-(N+1)^2\vert\bi\vert^2\right).\label{cnot}\ee
Using the notation
\eqref{hm} for $\lm=1$, the second-order action \eqref{cnot} can be rewritten as
 \be S(\vz)=\int dt\left(\vert \dot \vz\vert^2-\vert \vz^\dagger\dot \vz\vert^2+(N+1)^2(\vert Z_{N+1}\vert^2-1)+\Lambda(\vert\vz\vert^2-1) \right),\label{vz}\ee
   where $\Lambda $ is a real Lagrange multiplier and the complex $(N+1)$-vector $\vz$ has the components $$\vz=\bpm Z_1\\Z_2\\.\\.\\.\\Z_N\\Z_{N+1}\epm.$$ 
 The action \eqref{vz} is defined globally, that is it makes sense not only on the (dense open) chart $\bi$ but on the whole $\cpn$ space parametrized by the homogeneous coordinates normalized by the condition $\vert\vz\vert^2=1.$ It
   has a gauge symmetry
   $$ \vz\to e^{\ri\bt(t)}\vz,\quad \Lambda\to\Lambda-\ri\dot\bt(\vz^\dg\dot\vz-\dot\vz^\dg\vz)+\dot\bt^2\vert\vz\vert^2,$$
   where $\bt(t)$ is an arbitrary real function of the time $t$. On our chart where the coordinate $Z_{N+1}$ does not vanish, this gauge symmetry can be gauge fixed by claiming
   that $Z_{N+1}$ is real which is indeed solved by
   \be\vz=\bpm \chi_1\\\chi_2\\.\\.\\.\\\chi_N\\\sqrt{1-\vert\bi\vert^2}\epm\equiv \bpm\bi\\\ \\ \sqrt{1-\vert\bi\vert^2} \epm, \quad \vert\bi\vert<1.\label{gf}\ee 
   Inserting \eqref{gf} into the action \eqref{vz}, we recover the action \eqref{cnot}.
   
   \medskip 
\rem{\small  The standard Fubini-Study metric on the $\cpn$ expressed in  the normalized homogeneous coordinates is
   \be ds^2_{FS}=\vert d\vz\vert^2-\vert \vz^\dagger d\vz\vert^2.\label{fs}
   \ee
   We thus observe, that that our point particle principal chiral model corresponds to the motion in the standard Fubini-Study metric and in the quadratic potential depending on just  one of the normalized homogeneous coordinates.
   }

   \medskip

 We can make explicit also the first order action \eqref{fopc} . To do this,  we parametrize $\rho'\in\K_\zeta^\perp$ in the block matrix way as
    \be \rho'=\ri\bpm\boldsymbol{0}&\bw\\\bw^\dagger &0\epm,\label{rw}\ee  where $\bw$ is a complex $N$-vector. The first order action \eqref{fopc} on the chart where $Z_{N+1}$ does not vanish is then
   \be S(\bw,\bi)=-\int dt \Bigl(2\Im(\bw^\dagger(\dot\bi-\dot{\sq}\bi-\al(\bi^\dagger\dot\bi)\bi))  +\bw^\dg\bw+(N+1)^2\bi^\dg\bi\Bigr).\label{fopcbis}\ee
It is convenient to trade\footnote{{ Inversely, we have 
\be   \bw=\bp+\frac{\bp^\dg\bi-\bi^\dg\bp}{2\sqrt{1-\vert\bi\vert^2}}\bi+\frac{\al}{\sqrt{1-\vert\bi\vert^2}}(\bi^\dg\bp)\bi.\label{wp}\ee}} the variable $\bw$ for another one denoted as $\bp$
$$ \bp=\bw-\frac{\bw^\dg\bi-\bi^\dg\bw}{2\sqrt{1-\vert\bi\vert^2}}\bi-\al(\bi^\dg\bw)\bi, $$
the action \eqref{fopcbis} can be then rewritten as
 \be S(\bp,\bi)=\int dt\left(\ri\bp^\dg\dot\bi-\ri\dot\bi^\dg\bp - \bp^\dg\bp-\frac{(\bi^\dg\bp+\bp^\dg\bi)^2}{4(1-\vert\bi\vert^2)}-\js(\bi^\dg\bp-\bp^\dg\bi)^2-(N+1)^2\bi^\dg\bi\right).\label{fopctris}\ee

     Looking at the
     first order actions \eqref{fopcbis}, \eqref{fopctris},
     we can deduce the symplectic form and the Hamiltonian of the model to be
 \be \omega=-2d\Im(\bw^\dagger(d\bi-d({\sq})\bi-\al(\bi^\dagger d\bi)\bi)), \label{omn}  \ee\be H=\bw^\dg\bw+(N+1)^2\bi^\dg\bi,\label{han}\ee
or
 \be \omega=\ri d\bp^\dg \wedge d\chi+\ri d\bi^\dg\wedge  d\bp, \label{omnbis}  \ee\be H=\bp^\dg\bp+\frac{(\bi^\dg\bp+\bp^\dg\bi)^2}{4(1-\vert\bi\vert^2)}+\js(\bi^\dg\bp-\bp^\dg\bi)^2+(N+1)^2\bi^\dg\bi.\label{hanbis}\ee 
As for the first order equations of motion, they read
 \be \ri\dot\bi=\bp+\frac{\bi^\dg\bp+\bp^\dg\bi}{2(1-\vert\bi\vert^2)}\bi-\jp(\bi^\dg\bp-\bp^\dg\bi)\bi,\label{ceq1}\ee
\be \ri\dot\bp=\frac{\bi^\dg\bp+\bp^\dg\bi}{2(1-\vert\bi\vert^2)}\bp+(N+1)^2\bi+\frac{(\bi^\dg\bp+\bp^\dg\bi)^2}{4(1-\vert\bi\vert^2)^2}\bi+\jp(\bi^\dg\bp-\bp^\dg\bi)\bp.\label{ceq2}\ee}
{ \subsection{Symplectic reduction}
It is more convenient and also more elegant to rewrite the first order data \eqref{omn},\eqref{han} or \eqref{omnbis},\eqref{hanbis} in a global way, that is without restricting to the chart where $Z_{N+1}$ does not vanish. We do it via a suitable symplectic reduction of 
    a certain dynamical system $(\tilde\om,\tilde H)$ living on a complex symplectic      manifold
$\bc^{N+1}\times\bc^{N+1}$ parametrized by two
   $(N+1)$-vectors $\vy$, $\vz$,
$$ \vz=\bpm Z_1\\Z_2\\.\\.\\.\\Z_N\\Z_{N+1}\epm, \quad  \vy=\bpm Y_1\\Y_2\\.\\.\\.\\Y_N\\Y_{N+1}\epm.$$
 The (non-reduced) symplectic form on $\bc^{N+1}\times\bc^{N+1}$ is given by
 {  \be \tilde\om=\jp(d\vy^\dg\w d\vz-d\vz^\dg\w d\vy)=d\Re(\vy^\dg d\vz), \label{tom}\ee
   where
 $$ \vz^\dg=\bpm \bar Z_1 & \bar Z_2 & \dots &\bar Z_N &\bar Z_{N+1}\epm, \quad \vy^\dg=\bpm \bar Y_1 & \bar Y_2 & \dots &\bar Y_N &\bar Y_{N+1}\epm $$
  and  the Hamiltonian is given by\footnote{A similar Hamiltonian, however without the potential term, was recently considered in \cite{BS} as a basis of (quantum) symplectic reduction leading to the standard mechanical
$\cp^N$-model.}:  
 \be \tilde H(\vy,\vz)= \frac{1}{4}\vert\vz\vert^2\vert \vy\vert^2-\frac{1}{4}\vert \vz^\dg\vy\vert^2+(N+1)^2(1-\vert Z_{N+1}\vert^2).\label{tih}\ee

 Consider now a symplectic reduction of the dynamical system $(\tilde\om,\tilde H)$   induced by two Poisson-commuting constraints 
    \be  \vy^\dg\vz-\vz^\dg\vy=0,\quad \vz^\dg\vz-1=0,\label{con}\ee 
   which generate, via the Poisson brackets, two commuting flows 
 \be \phi_{\bt_1}(\vy,\vz)=(e^{\ri \bt_1}\vy,e^{\ri \bt_1}\vz),\quad \phi_{\bt_2}(\vy,\vz)=(\vy+\bt_2\vz,\vz) .\label{orb}\ee
   It turns out, that this symplectic reduction  gives precisely the first order data \eqref{omn},\eqref{han}  of the point particle principal chiral
   model \eqref{vz}. 
   Indeed, on the chart where $Z_{N+1}$ does not vanish, we may completely slice
   the flows \eqref{orb} on the constraints \eqref{con} by the following supplementary constraints  
   \be  Z_{N+1}-\bar Z_{N+1}=0,\quad \vy^\dg\vz+\vz^\dg\vy=0 .\label{adc}\ee
   In order to describe the symplectic form $\tilde\om$ restricted to the submanifold \eqref{con},\eqref{adc}, we first solve the constraints \eqref{con},\eqref{adc} in terms of two $N$-vector variables 
 $\vxi,\bi$   such that $\vert\bi\vert<1$:
 { \be \vy=-2\ri\bpm \vxi\\-\frac{\bi^\dg\vxi}{\sqrt{1-\bi^2}}\epm, \quad\vz=\bpm\bi\\\ \\\sqrt{1-\bi^2} \epm,\label{soc}\ee}

Set also\footnote{{ Inversely, we have
\be  \xi=\bw-\al(\bi^\dg\bw)\bi,\label{xiw}\ee
and we have also 
$$ \bp=\bxi+\frac{\bi^\dg\bxi-\bxi^\dg\bi}{2(1-\vert\bi\vert^2)}\bi, \quad \xi=\bp-\jp(\bi^\dg\bp-\bp^\dg\bi)\bi,$$
 $$ \bi^\dg\bp-\bp^\dg\bi=\frac{\bi^\dg\bw-\bw^\dg\bi}{\sqrt{1-\vert\bi\vert^2}}=\frac{\bi^\dg\bxi-\bxi^\dg\bi}{ {1-\vert\bi\vert^2}}, \quad \bi^\dg\bp+\bp^\dg\bi=\sqrt{1-\vert\bi\vert^2}(\bi^\dg\bw+\bw^\dg\bi)= \bi^\dg\bxi+\bxi^\dg\bi . $$}}
{ \be \bw:=\sqrt{\1_N+\frac{\bi\otimes\bi^\dg}{1-\vert\bi\vert^2}}\vxi=\vxi+\frac{\al}{\sqrt{1-\vert\bi\vert^2}}(\bi^\dg\vxi)\bi.\label{wxi}\ee}
 With the notation \eqref{wxi},   the symplectic form $\tilde\om$ and the Hamiltonian $\tilde H $ restricted to \eqref{con},\eqref{adc} become, respectively,
 { $$\omega=-2d\Im(\bw^\dagger(d\bi-d({\sq})\bi-\al(\bi^\dagger d\bi)\bi)),$$
$$  H(\bw,\bi)=\bw^\dg\bw+(N+1)^2\bi^\dg\bi.$$}
We have thus recovered the principal chiral first order data \eqref{omn},\eqref{han}.

\medskip

  \medskip

We wish also to rewrite the first order equations of motion \eqref{ceq1},\eqref{ceq2} in the global way, that is without restriction to  the chart where $Z_{N+1}\neq 0$. To do that we first derive the  equations of motion of the non-reduced dynamical system  $(\tilde\om,\tilde H)$.
 We deduce from \eqref{tom},\eqref{tih}
     that the first order action of the non-reduced system $(\tilde\om,\tilde H)$   reads \be S(\vy,\vz)=\int dt (\Re(\vy^\dg\dot\vz)-\tilde H(\vy,\vz))\label{hfom}.\ee
   By varying the action \eqref{hfom}, we find 
  \be \dot\vz-\jp\vy\vert \vz\vert^2+\jp (\vz^\dg\vy)\vz=0,\label{z}\ee
  \be \dot\vy+\jp\vert  \vy\vert^2\vz-\jp (\vy^\dg\vz)\vy-2(N+1)^2  Z_{N+1}\en=0.\label{y}\ee

  \rem{\small 
  Alternatively, we can find the non-reduced equations of motion \eqref{z},\eqref{y} by calculating the Poisson brackets of $\vz,\vy$ with $\tilde H$, knowing that the Poisson brackets induced by $\tilde\om$ are
$$\{Z_j,\bar Y_k\}=\{\bar Z_j,Y_k\}=2\delta_{jk}, \quad\{Z_j,\bar Z_k\}=\{Y_j,\bar Y_k\}=0,$$
   $$ \{Z_j,Y_k\}=\{Z_j,Z_k\}=\{Y_j, Y_k\}=\{\bar Z_j,\bar Y_k\}=\{\bar Z_j,\bar Z_k\}=\{\bar Y_j, \bar Y_k\}=0.$$}

  \rem{\small  Because the constraints \eqref{con} Poisson-commute with the unreduced Hamiltonian $\tilde H$,   we have also { \be \frac{d(\vz^\dg\vz-1)}{dt}=0,\quad\frac{d(\vz^\dg\vy-\vy^\dg\vz)}{dt}=0.\label{vc}\ee}  Of course, the validity of \eqref{vc} can be established also as the direct consequence  of \eqref{z}, \eqref{y}.}
\medskip

 By the general principles of symplectic reduction, the reduced dynamics is nothing but the unreduced dynamics of flow-invariant observables on the constraint surface \eqref{con}.  
Setting
\be { \vk:=\vy-(\vz^\dg\vy)\vz,}\label{ky}\ee
 we observe that the variables $\vz,\vk$ are flow invariant with respect to the second flow $\phi_{\bt_2}$ but not with respect to the first one $\phi_{\bt_1}$. However, the following matrix variables $\W,\J$
 are flow invariant with respect to the both flows
  \be \W:=\ri(N+1)(\vz\otimes\vz^\dg-\en\otimes\en^\dg),\quad \J=\jp\vk\ot\vz^\dg- \jp\vz\otimes \vk^\dg.\label{JF}\ee
  Note that both $\W$ and $\J$ are traceless anti-Hermitian matrices.

  \medskip
  
From \eqref{z}, \eqref{y}, we find that  on the constraint surface \eqref{con} it holds

 $$ \dot\vz=\jp\vk,  \quad \dot\vk=-\jp\vert\vk\vert^2\vz+2(N+1)^2Z_{N+1}(\en-\bar Z_{N+1}\vz),\label{zbis} $$
 and hence

 \be\dot\W  =[\J,\W-\zeta],\quad \dot\J= [\zeta,\W],\label{ab} \ee
where $\zeta$ is given by \eqref{cpz}.
The equations \eqref{ab} are the reduced first order equations of motions written in the global way.

\rem{\small It is instructive to see how we can recover form \eqref{ab} the first order equations of motions \eqref{ceq1},\eqref{ceq2} valid  on the chart where $Z_{N+1}\neq 0$.  For that, we use Eqs.\eqref{wp},\eqref{soc},\eqref{wxi},\eqref{ky},\eqref{JF} to express the components of the $N$-vector variables
$\bi,\bp$ in terms of the matrix elements of the matrices $\J,\W$. For $\al=1,\dots,N$, we find 
  \be \chi_\al=\frac{-\ri\W_{\al,N+1}}{\sqrt{(N+1)(-\ri\W_{N+1,N+1}+N+1)}},\label{cpa}\ee 
  \be p_\al=\ri\frac{(N+1)\J_{\al,N+1}-\ri[\J,\W-\zeta]_{\al,N+1}}{2\sqrt{(N+1)(-\ri\W_{N+1,N+1}+N+1)}}+\ri\frac{\W_{\al,N+1}[\J,\W-\zeta]_{N+1,N+1}}{2\sqrt{N+1}(-\ri\W_{N+1,N+1}+N+1)^{\frac{3}{2}}}.\label{cpb}\ee
We note, that the expressions on the r.h.s. of \eqref{cpa},\eqref{cpb} are well defined on the chart where $Z_{N+1}\neq 0$. Finally, we use \eqref{ab} to calculate the time derivatives of   \eqref{cpa},\eqref{cpb} which gives, after some work, precisely \eqref{ceq1},\eqref{ceq2}.}

\medskip

Define now two $\lm$-dependent matrices $L(\lm),M(\lm)$ as
   \be L(\lm)= \zeta - \frac{\lm\J+\W}{1-\lm^2},\quad \ M(\lm)= -\frac{\J+\lm \W }{1-\lm^2},\label{lpc}\ee 
   where $\zeta$ is given by \eqref{cpz}.
   Inserting the expressions
   \eqref{lpc} into the Lax equation $\dot L(\lm)=[L(\lm),M(\lm)]$ gives
   \be -\frac{\lm\dot\J+\dot\W}{1-\lm^2}=\frac{-\lm[\zeta,\W]+[\W-\zeta,\J]}{1-\lm^2}.\label{lpd}\ee
  The validity of the relation \eqref{lpd} for every $\lm$ is thus equivalent to the fulfillment of the following two conditions
  $$\dot\W  =[\J,\W-\zeta],\quad \dot\J= [\zeta,\W] , $$
   which are nothing but the reduced equations of motion \eqref{ab}.
 In this way, we have just established the Lax integrability of the point particle principal chiral model on $\cpn$.
     It may seem that we have just pulled out the Lax pair \eqref{lpc}  out  of a hat but it is not the case. Actually, by the general construction presented in Section 5.2 we know that the principal chiral
     model admits the Lax pair \eqref{laxp}, but what is the relation between \eqref{laxp} and \eqref{lpc}? Well, it turns out that 
     the Lax pair \eqref{laxp} is nothing but the
     Lax pair \eqref{lpc}. This can be seen by inserting the formulas \eqref{kchi},\eqref{rw} into \eqref{laxp} and using the formulas \eqref{soc},\eqref{xiw}.} In particular, it holds
$$ \J=-k\rho'k^{-1},\quad \W=k'k^{-1}, \quad l'l^{-1}=\left( \W,-\J\right). $$

\section{Lu-Weinstein Drinfeld double $SL(N+1,\bc)$}

\subsection{Bi-Yang-Baxter deformation: the general case}

{ In this section, we work out the point particle version of a particular two-parametric deformation of the principal chiral model which is called in the literature the bi-Yang-Baxter model or the Klim\v c\'ik  model \cite{K02,K08}. It is a particular case of the  $\E$-model based on the Lu-Weinstein
    Drinfeld double $D$, where  $D$ is
    the special complex linear group ${ SL(N+1,\bc)}$ viewed as {\it real} group (i.e. it has the dimension $2(N^2+2N)$ as the real manifold). 
The  non-degenerate symmetric ad-invariant bilinear form ${ (.,.)_{\D}}$ on the Lie algebra ${sl(N+1,\bc)}$ is defined by taking a suitable normalized  imaginary
 part of the trace
 $$ {(X,Y)_{\D}= \frac{-1}{\eta}\Im\tr(XY), \quad X,Y\in  sl(N+1,\bc)}, \quad \eta>0.$$
 The two half-dimensional isotropic  subgroups $K$ and $ \tilde K$ are
 respectively the special unitary group ${ SU(N+1)}$ and  the upper-triangular group ${ AN}$ with real positive numbers on the diagonal the product of which is equal to $1$.
 
 \medskip
 We choose for $\xi$ an element of the Cartan subalgebra of the isotropic subalgebra  $\K$. The symplectic form then reads
   $$ \omega^\xi=-d(\xi, l^{-1}dl)_\D=\frac{1}{\eta}{d\Im\tr(\xi  l^{-1}dl)}
  $$
 The two-parametric family of the $\E$-operators, is given by \cite{K16}
 $$ \E_{\eta,\mu} X=\frac{\ri}{2} \frac{\eta^2+\mu^2R^2-1}{\eta}X-\frac{\ri}{2} \frac{\eta^2-\mu^2R^2+1}{\eta}X^\dagger- \mu R X^\dagger, \quad X\in sl(N+1,\bc),$$
where  $R:sl(N+1,\bc)\to sl(N+1,\bc)$ is the Yang-Baxter operator  defined  on the Chevalley basis   of   $sl(N+1,\bc)$ as
\be RE^{\al}=-{\rm sign}(\al){\rm i}E^{\al},\quad RH^j=0.\label{ybr}\ee

\medskip

  The Hamiltonian of the point particle bi-Yang-Baxter  model  then reads
$$ H^\xi_{\eta,\mu}  =\jp \bigl(l'l^{-1},\E_{\eta,\mu} \ \! l'l^{-1}\bigr)_{\D}=\frac{\mu}{2\eta}\Im\tr\left(l'l^{-1}R(l'l^{-1})^\dagger\right)+$$\be +
 \js \tr\left(l'l^{-1}(\eta^{-2}-\mu^2\eta^{-2}R^2+1)(l'l^{-1})^\dagger\right)+\js  \Re\tr\left(l'l^{-1}(\eta^{-2}-\mu^2\eta^{-2}R^2-1)(l'l^{-1})\right)  .\label{xiHbr}\ee}

 \medskip
  
{ We may parametrize $l\in D$  in the Iwasawa way as
 \be l=gb=gan,\quad g\in SU(N+1),\ b=an\in AN, \quad a\in A,\ n\in N.\label{iwr}\ee
 We have denoted by $A$
   the subgroup of the diagonal matrices of $AN$
 and by $N$   the subgroup of the matrices in $AN$ having just units on the diagonal.

 \medskip
 
 In a dense open subset of $SU(N+1)$, we can parametrize $g$  as 
  $$ g=kk_\xi, $$
 where $k_\xi$ belong to the stabilizer $K_\xi$.  In this parametrization, the symplectic form $\om^\xi$
 becomes
$$ \om^\xi=-d(\xi,l^{-1}dl)_\D =-d(b\xi b^{-1}, k_\xi^{-1}k^{-1}dk k_\xi+k_\xi^{-1}dk_\xi)_\D-d(\xi,n^{-1}a^{-1}dan+n^{-1}dn )_\D=$$
\be =-d(k_\xi b\xi b^{-1}k_\xi^{-1}, k^{-1}dk)_\D-d(\xi,n^{-1}a^{-1}dan)_\D=-d(k_\xi b\xi b^{-1}k_\xi^{-1}-\xi, k^{-1}dk)_\D.\label{symqr}\ee
We find in the block form
 $$ k_\xi b\xi b^{-1}k_\xi^{-1}-\xi=-\eta(R-\ri)[\xi,\rho],$$
where $\rho\in\K$  and $R$ is the Yang-Baxter operator.
\medskip

As far as the Hamiltonian is concerned, we first use the Iwasawa decomposition \eqref{iwr} to find
$$ l'l^{-1}=k'k^{-1}+\eta(R_{k^{-1}}-\ri)(k\rho'k^{-1}), $$
where
$$ R_{k^{-1}}:=\Ad_{k}R\Ad_{k^{-1}}.$$
Then we use \eqref{xiHbr} to find
\be H^\xi_{\eta,\mu}=-\jp\tr\left(   (\rho')^2 +(k^{-1}k'+(\eta R+\mu R_k)\rho')^2\right).\label{hanqr}\ee

\medskip

The first order action corresponding to the data \eqref{symqr},\eqref{hanqr} is therefore
 \be S_{\eta,\mu}(k,\rho')=\int dt\  \tr\Bigl( \rho'k^{-1}\dot k +\jp   (\rho')^2 +\jp(k^{-1}k'+(\eta R+\mu R_k)\rho')^2 \Bigr).\label{fopcer}\ee}

\medskip

 Note that the variable $\rho'$ takes values\footnote{Note also that the Yang-Baxter operator $R$ restricted to the subspace $K^\perp_\xi$ has  a trivial kernel, which means that it behaves as a  complex structure similarly as in the infinite dimensional context described in \cite{Byk}.} in $K^\perp_\xi$. It is sometimes convenient to use a variable $\chi\in\K$ instead of $\K_\xi$ and a Lagrange multiplier $A\in \K_\xi$ forcing $\chi$ to become $\rho'\in \K^\perp_\xi$. Explicitely,
 $$  S_{\eta,\mu}(k,\chi,A)=\int dt\  \tr\Bigl( \chi (k^{-1}\dot k -A)+\jp\chi^2+\jp(k^{-1}k'+(\eta R+\mu R_k)\chi)^2\Bigr). $$ 
We can then easily eliminate the field $\chi$, which gives
    $$ S_{\eta,\mu}(k,A)=
   -\int \ dt \ \tr\left((k^{-1}\dot k-A+k^{-1}k')\frac{1}{1-\eta R-\mu R_k}(k^{-1}\dot k-A-k^{-1}k')\right),$$
   or, equivalently,
       \be S_{\eta,\mu}(k,A)=
   \int \ dt \ \tr\left((k^{-1}k')^2- (V(k)-A)\frac{1}{1-(\eta R+\mu R_k)^2}(V(k)-A) \right),\label{eqaub}\ee
where    $$ V(k):= k^{-1}\dot k -(\eta R +\mu R_k)k^{-1}k'. $$

   \medskip
      
  Varying the actions \eqref{fopcer} and \eqref{eqaub} respectively with respect to $\rho'$ and $A$ gives
  \be \rho'(k)=-\frac{1}{1_\perp-P^\perp(\eta R+\mu R_k)^2P^\perp}P^\perp(k^{-1}\dot k -(\eta R +\mu R_k)k^{-1}k'),\label{skeer}\ee
  \be A(k)=PV(k)+P(\eta R+\mu R_k)^2P^\perp \frac{1}{1_\perp-P^\perp(\eta R+\mu R_k)^2P^\perp}P^\perp V(k),\label{ak}\ee
  where $P$ and $P^\perp$ are orthogonal projectors on $K_\xi$ and $K_\xi^\perp$, respectively.

  \medskip 

Inserting \eqref{skeer} back into
  the action \eqref{fopcer} or \eqref{ak}  back in \eqref{eqaub} gives rise to the second order action
   \be S_{\eta,\mu}(k)=\jp\int dt\ \tr\left(-P^\perp V(k) \frac{1}{1_\perp-P^\perp(\eta R+\mu R_k)^2P^\perp} P^\perp V(k)+(k^{-1}k')^2\right) .\label{sor}\ee
 
  By construction, the action \eqref{sor} lives on the flag\footnote{In  \cite{B20},  r-matrix-deformations of   flag
manifold $\si$-models were studied. The standard dimensional reduction of the models \cite{B20} gives  
mechanical models  which differ from ours because they miss potential terms and only the metric is deformed.
As we have already stated in the Introduction, this is consistent because what we do here is not quite
the dimensional reduction. Indeed,  we do not let the space derivative 
terms vanish but we rather replace them with the commutators with
$\xi$ which yields indeed the potential terms. The "raison d'\^etre" of our procedure resides in the fact that we obtain in this way the mechanical  Lax pairs 
from the field-theoretical ones automatically. Whether 
 Lax pairs (if they exist)  of the standardly dimensionally reduced models \cite{B20} could be also  obtained  in one or in other way from the field theoretical ones  is not clear to us.}  manifold $K/K_\xi$. This can be verified directly by 
checking its  gauge invariance   with respect to the gauge transformation      $k\to kh$, $h\in K_\xi$. First of all we find
 $$ V(kh)=\Ad_{h^{-1}}  (V(k)-\eta\Pi_h k^{-1}k')
 +h^{-1}\dot h, $$
  $$ P^\perp(\eta R+\mu R_{kh})^2P^\perp=P^\perp(\eta R+\mu \Ad_{h^{-1}}R_{k}\Ad_h)^2P^\perp=P^\perp\Ad_{h^{-1}}(\eta R+\mu R_k+\eta \Pi_h)^2 \Ad_hP^\perp, $$
  where
  $$ \Pi_h:=(R_{h^{-1}}-R). $$
  Because of the relations
   $$[R,P^\perp]=0,\quad [\Ad_h,P^\perp]=0,\quad [R,\Ad_h P^\perp]=0, $$   we infer that the image of the operator $\Pi_h$ is in $\K_\xi$, while $K^\perp_\xi$ is in the kernel of $\Pi_h$. This implies
  \be P^\perp V(kh)=\Ad_{h^{-1}} (P^\perp V(k)),\label{gv}\ee
\be P^\perp(\eta R+\mu R_{kh})^2P^\perp= \Ad_{h^{-1}}P^\perp(\eta R+\mu R_k+\eta \Pi_h)^2P^\perp\Ad_h =\Ad_{h^{-1}}P^\perp(\eta R+\mu R_k )^2P^\perp\Ad_h. \label{pwpb}\ee
 The relations \eqref{gv}, \eqref{pwpb} then
 imply the gauge invariance of the action
 \eqref{sor}.}
\subsection{Integrability}
   Let us show that the point particle bi-Yang-Baxter  model is integrable for whatever choice of $\xi$. For that we  take  the Lie algebra $\G$ to be just $\K$ equipped with its standard Killing-Cartan form $(.,.)_\G=(.,.)_\K$ and, by the very logic of this article, we borrow  the required operators $O(\lm), \hat r(\lm,\rho)$ from the stringy bi-Yang-Baxter context. They  are given by \cite{K20}
 \be O(\lm)j=f_0(\lm)\frac{j-j^\dagger}{2}+\left(  f_1(\lm)+\mu (1-f_0(\lm))R   \right)\frac{j+j^\dagger}{2 \ri\eta},\label{obi}\ee
  $$ O(\lm)^\dagger W=(-\eta \ri f_0(\lm)- f_1(\lm)+\mu(1-f_0(\lm))R)W,$$
 \be \hat r_{\rm bi-YB}(\lm,\rho)= (1-f_0(\rho)) 
\left(\frac{f_0(\lm)f_1(\rho)+f_1(\lm) f_0(\rho)}{f_0(\lm)-f_0(\rho)}{\rm Id}-\mu R\right),\label{rbi}\ee
 where
 $$f_0(\lm)=\frac{1+\eta^2-\mu^2}{2}+\frac{\sqrt{(1+\eta^2-\mu^2)^2+4\mu^2}}{2}\cosh{\lm},$$
 $$f_1(\lm)= \frac{\sqrt{(1+\eta^2-\mu^2)^2+4\mu^2}}{2}\sinh{\lm}.$$


 We know that $O(\lm),\hat r(\lm,\rho)$ given by \eqref{obi}, \eqref{rbi} satisfy the stringy conditions
 \eqref{sufcb},  \eqref{1db}, \eqref{2db}  automatically. We must then check  that the purely point particle  conditions \eqref{sucb}, \eqref{suc2b} are also satisfied.
 This follows from  the fact that the 
 Yang-Baxter operator $R$ commutes with $\ad_\xi$, as can be checked from \eqref{ybr} and from the relation
 $$\ad_\xi E^\al=\al(\xi)E^\al, \quad \ad_\xi H^j=0.$$

We conclude, by the general construction of Section 4, that the Lax pair of the point particle bi-Yang-Baxter model is therefore given by \eqref{Laxpp}, \eqref{ppM}
$$    L(\lm)  = \xi- O(\lm)j, \quad 
   M(\lm)=-  O(\lm)\E j.$$  
 
\medskip
{ \subsection{Bi-Yang-Baxter deformation: particular cases}
 Let us evaluate the second order action \eqref{sor} for the group $K=SU(2)$. In this case, the Yang-Baxter operators $R$, $R_k$ can be written as
   $$ R=-\frac{1}{2}\ad_\xi, \quad  R_k=R-\jp\ad_{k^{-1}k'}. $$
  Using  the parametrization
   $$ k=\bpm u&-\bar v\cr v&\bar u\epm, \quad u\bar u+v\bar v=1, $$
 we find
 $$ P^\perp V(k)=\bpm 0 &\bar vd\bar u-\bar u d\bar v+2(\eta+\mu)\bar u\bar v\cr udv-vdu-2(\eta+\mu)uv&0\epm. $$
Setting
 $$ x_1-\ri x_2=2u\bar v, \quad x_3=u\bar u-v\bar v,$$
 we obtain
$$\left(1_\perp-P^\perp(\eta R+\mu R_k)^2P^\perp\right)^{-1} \ri\bpm 0& y_1-\ri y_2\cr y_1+\ri y_2 &0 \epm=$$ $$ =\ri(1+\eta^2+\mu^2+2\mu\eta x_3)^{-1}\left(\bpm 0& y_1-\ri y_2\cr y_1+\ri y_2 &0 \epm +\frac{4\mu^2 \Re{(uv(y_1-\ri y_2))}}{1+(\eta+\mu x_3)^2}\bpm 0& \bar u\bar v\cr uv &0 \epm\right). $$}

 The action \eqref{sor} thus becomes
{     \be S_{\eta,\mu}=\frac{1}{4}\int dt \left(\frac{\dot x_1^2+\dot x_2^2+\dot x_3^2+\frac{\mu^2(x_1\dot x_2-x_2\dot x_1)^2}{1+(\eta+\mu x_3)^2}}{{1+\eta^2+\mu^2+2\mu\eta x_3}}+4(x_3-1)\frac{2-(\eta-\mu)^2(x_3-1)  }{1+\eta^2+\mu^2+2\mu\eta x_3}\right)\label{sor2}\ee

Working in the standard spherical coordinates
 $$ x_1=\sin{\theta}\cos{\phi},\quad x_2=\sin{\theta}\sin{\phi},\quad x_3=\cos{\theta}, $$
the action \eqref{sor2} becomes
$$ S_{\eta,\mu}=\frac{1}{4}\int dt 
\left(\frac{\dot\theta^2}{1+\eta^2+\mu^2+2\mu\eta \cos{\theta}}+\frac{\sin^2{\theta}\dot\phi^2}{1+\eta^2+\mu^2\cos^2{\theta}+2\mu\eta \cos{\theta}}\right)+$$ $$+
\int dt (\cos{\theta}-1)\frac{2-(\eta-\mu)^2(\cos{\theta}-1)}{1+\eta^2+\mu^2+2\mu\eta \cos{\theta}}. $$

\medskip 

For $\mu=0$, we have

     $$ S_{\eta}=\frac{1}{4}\int dt \left(\frac{\dot x_1^2+\dot x_2^2+\dot x_3^2 }{1+\eta^2}+4(x_3-1)\frac{2-\eta^2(x_3-1)  }{1+\eta^2}\right)$$
     $$  S_{\eta}=\frac{1}{4}\int dt \left(
\frac{\dot\theta^2+\sin^2{\theta}\dot\phi^2}{1+\eta^2}+ 
4 (\cos{\theta}-1)\frac{2-\eta^2(\cos{\theta}-1)}{1+\eta^2}\right).$$

We note  that,  up to a renormalization by a constant,   the Fubini-Study metric does not change in the case $\mu=0$ and only the potential term  gets substantially deformed.}

\medskip 

  Consider now the
case $K=SU(3)$ and
 $$\xi=\ri\bpm 1 &0&0\\0&1&0\cr 0&0&-2\epm. $$
We give the details of  the computation in Appendix, here we write just the result  in the following dense parametrization \cite{By} of the homogeneous space
$K/K_\xi=\cp^2$

  $$ k=\bpm a&-\bar b&0\cr b&\bar a&0\cr 0&0&1\epm\bpm \cos{\theta}&0&\sin{\theta}\cr 0&1&0\cr -\sin{\th}&0&\cos{\th}\epm, \quad a\bar a+b\bar b=1. $$
The bi-YB action \eqref{sor} then is
$$    S_{\eta,\mu}(k)= \int dt\   \Delta_2\left(\frac{\bt}{\vert a\vert^2\vert b\vert ^2} W_5^2+\e_2 W_6^2+2qW_5W_6\right)+$$
\be + \int dt \   \Delta_1\left(\left(\frac{\bt}{\vert a\vert^2\vert b\vert ^2}+\delta\right)W_4^2+\e_1 W_7^2+2pW_4W_7\right) -9\int dt \sin^2{\th}  , \label{3so}\ee
where
 $$\  W_4=\st\ct\ \Im(\bar a\dot a+\bar b\dot b), \quad W_5=\dot\th-3\st\ct(\eta+\mu),$$ $$  W_6= \st\Re\left(\frac{\dot  b }{b}-\frac{\dot{ a}}{a}\right)+3\mu \sin^3{\th},\quad W_7=-\st\Im\left(\frac{\dot  b }{b}-\frac{\dot{ a}}{a}\right),  $$
$$\e_1=1+(\eta+\mu\cos{2\th})^2+\mu^2\sin^2{2\th}\vert a\vert^2\vert b\vert^2,\ \e_2=1+\eta^2+\mu^2+2\eta\mu\cos{2\th},\  \delta=-4\mu^2\sin^2{\th}, $$
$$   p=\mu\sin{2\th} \st(\eta-\mu(\vert a\vert^2-\vert b\vert^2)), \  q=\eta\mu \sin{2\th} \st, \  \beta=1+(\eta+\mu)^2-4\eta\mu \sin^2{\th}\vert a\vert^2,$$
$$\Delta_1=\frac{\vert a\vert^2\vert b\vert ^2}{\e_1\bt+(\e_1\delta-p^2)\vert a\vert^2\vert b\vert ^2 }, \quad \Delta_2=\frac{\vert a\vert^2\vert b\vert ^2}{\e_2\bt -q^2\vert a\vert^2\vert b\vert ^2 }.$$
Similarly as in the $SU(2)$ case, the second order action \eqref{3so} describes a dynamical system living on the complex projective space  equipped with the doubly deformed Fubini-Study metric and with a nontrivial potential. Needless to say, the action
\eqref{3so} is somewhat cumbersome, however, it is quite remarkable that it is underlied by the simple and transparent first order Hamiltonian structure given by the point particle $\E$-model.

\medskip 

Finally, we work out the case $K=SU(N)$, $\mu=0$ for $\xi$ given by the block matrix
 $$\xi=\ri\bpm \1_N&0\\ 0&-N\epm. $$
For that, we parametrize $\cp^N$ like in \eqref{kchi}
    {  $$ k=\bpm \1_N-\al \bi\otimes \bi^\dg & \bi\\ -\bi^\dagger&\sqrt{1-\vert\bi\vert^2}\epm,\quad \alpha=\frac{1-\sqrt{1-\vert\bi\vert^2}}{\vert \bi\vert^2},\quad \bi\neq\boldsymbol{0}, $$}
    which gives, in particular
   {   $$ k^{-1}k'=(N+1)\ri\bpm -\bi\otimes\bi^\dagger &\sq\bi\\\sq\bi^\dagger&\vert\bi\vert^2\epm,
    $$
   $$ P^\perp k^{-1}\dot k=
    \bpm \boldsymbol 0&\dot\bi-\dot{\sq}\bi-\al(\bi^\dagger\dot\bi)\bi\\-\dot\bi^\dagger+\dot{\sq}\bi^\dagger+\al(\dot\bi^\dagger\bi)\bi^\dagger&0\epm $$
$$(k^{-1}k',k^{-1}k')_\K=-2(N+1)^2\vert\bi\vert^2. $$}
    The point particle Yang-Baxter model action \eqref{sor} for $\mu=0$ therefore becomes
    $$   S_\eta(\bi)=\jp\int dt\left(- (1+\eta^2)^{-1}\left(P^\perp (k^{-1}\dot k -\eta Rk^{-1}k')\right)^2+(k^{-1}k')^2\right)=$$ \be=\int dt\left(\frac{\left\vert \dot\bi-\dot{\sq}\bi-\al(\bi^\dagger\dot\bi)\bi-\eta(N+1)\sq\bi\right\vert^2}{1+\eta^2}-(N+1)^2\vert\bi\vert^2\right).\label{cnote}\ee

 \medskip

Using  the  homogeneous coordinates (cf. \eqref{gf}), we obtain a more elegant global expression for the Yang-Baxter  second order action \eqref{cnote}
$$ S_\eta(\vz)=$$ $$=\int dt\left(\frac{\vert \dot \vz\vert^2-\vert \vz^\dagger\dot \vz\vert^2}{1+\eta^2}+\frac{(N+1)^2}{1+\eta^2}\left((2\eta^2+1)\vert Z_{N+1}\vert^2
- \eta^2\vert Z_{N+1}\vert^4\right)+\Lambda(\vert\vz\vert^2-1) \right). $$
  
\section{Conclusions and perspectives}

In the present article, we have introduced the {\it point particle $\E$-models}, which are the first order Hamiltonian dynamical systems associated to a Drinfeld double $D$, to an element $\xi$ of the quadratic Lie algebra $\D$  of $D$ and to a symmetric involution $\E:\D\to\D$.  Comparing with the infinite dimensional  stringy $\E$-models, based solely  on the data $(D,\E)$, the point particle models are finite
dimensional; the dimensions of their phase spaces $D/D_\xi$ depend on the dimension of $D$ and of the stabilizer subgroup $D_\xi$.  

\medskip 

Similarly as in the string case, we have shown that many of the point particle $\E$-models are integrable and  we have also identified
the sufficient conditions  which ensure the integrability. Those conditions \eqref{sufco}, \eqref{1do}, \eqref{2do}  are in fact the same in the point particle and in the string case, which means that  it is   easy to construct many new integrable point particle models by borrowing from the string literature the known solutions of \eqref{sufco}, \eqref{1do}, \eqref{2do}.  The only   thing to check is  whether the chosen  $\xi\in\D$ verifies the supplementary purely point particle condition \eqref{suc}.   Fortunately, the condition \eqref{suc} turns out to be only mildly restrictive, in particular, for the Drinfeld doubles $T^*SU(N)$ and $SL(N,\bc)$ many choices of $\xi$ satisfy it,  leading to the
physical systems living on the complex flag manifolds.

\medskip

As for the perspectives, the obvious task would be to generalize the point particle $\E$-models   along the lines of the string generalization leading from the standard (or non-degenerate) $\E$-models
to the degenerate ones \cite{KS96b,K19}. If this point particle generalization can be found, the next question to ask would be what are the sufficient condition of its integrability. The fact that those sufficient conditions  are already known in the degenerate string case \cite{K22} should be helpful.

\medskip

Another open issue is a quantization  of the point particle $\E$-models, in particular of the integrable ones. This does not look a priori difficult, because, similarly as in the stringy case, the basic observables of the point particle $\E$-models are the $\D$-valued currents $j$ verifying the finite dimensional version of the Poisson current algebra \eqref{xicur}. Since the Poisson brackets \eqref{xicur} seem easy to quantize, this  should lead  to the complete quantization of the model provided we know well the theory of representations of the given Drinfeld double $D$. 

\medskip

A successful  quantization of the integrable point particle   $\E$-models could bring about also an additional benefit, namely, it could shed   light on the difficult problem of the quantization of the so-called non-ultralocal theories \cite{M1,M2,Ko19,KLT} among which typically belong the integrable non-linear $\sigma$-models. In the non-ultra-local case, the Lax matrix elements at different values of the loop parameter do not commute. There is no loop parameter in the finite dimensional case, so that the ultra-local problem cannot arise there, but the algebraic similarity of the finite and of the infinite, notably almost the same structure of the Lax operator and of the $r$-matrix, might provide  some clues how to quantize in the infinite case once the finite case quantization is  accomplished.

\medskip

In the infinite-dimensional case, the $\E$-models  are submitted to the renormalization group flow, that is a given $\E$-model flows in the into another one   and the direction and the rapidity of the flow is given by the initial value of the operator $\E$. In particular, there are algebraically distinguished "fixed point" or "conformal" operators $\E$ which do not flow at all. The detailed classification of those fixed points was recently performed in \cite{L23} in the both non-degenerate and degenerate cases. It could be interesting to work out the proporties of the conformal $\E$-models in the point particle case.

\section*{Appendix} 
Here we work out in  detail the bi-YB action \eqref{sor} for the
case $K=SU(3)$ and
 $$ \xi=\ri\bpm 1 &0&0\\0&1&0\cr 0&0&-2\epm. $$
The action lives on the homogeneous space
$K/K_\xi=\cp^2$ which we densely parametrize  as in  \cite{By} 

 $$ k=\bpm a&-\bar b&0\cr b&\bar a&0\cr 0&0&1\epm\bpm \cos{\theta}&0&\sin{\theta}\cr 0&1&0\cr -\sin{\th}&0&\cos{\th}\epm, \quad a\bar a+b\bar b=1. $$

 We then find
 $$ k^{-1}k'=3\ri\sin{\th}\bpm -\sin{\theta}&0&\cos{\theta}\cr 0&0&0\cr \cos{\th}&0&\sin{\th}\epm, \quad Rk^{-1}k'=3\sin{\th}\bpm 0&0&\cos{\theta}\cr 0&0&0\cr -\cos{\th}&0&0\epm, $$
 
  $$ k'k^{-1}=3\ri\sin{\th}\bpm a\bar a\sin{\theta}&a\bar b\st&a\cos{\theta}\cr b\bar a\st&b\bar b\st&  b\ct\cr \bar a\cos{\th}&\bar b\ct&-\sin{\th}\epm,$$ $$Rk'k^{-1}=3\sin{\th}\bpm0&a\bar b\st&a\cos{\theta}\cr -b\bar a\st&0& b\ct\cr - \bar a\cos{\th}&- \bar b\ct&0\epm,$$
 $$ R_k k^{-1}k'=-3\st\bpm 0&-\bar a\bar b\st\ct &-\ct\cr \st\ct ab&0&ab\sin^2{\th}\cr \ct &-\sin^2{\th}\bar a\bar b &0\epm, $$

  $$ k^{-1}\dot k=\bpm \cos^2{\th}(\bar a\dot a+\bar b\dot b)&\ct(\dot{\bar a}\bar b-\bar a\dot{\bar b})&\st\cos{\th}(\bar a\dot a+\bar b\dot b)+\dot\th\cr  \ct(a\dot b-\dot a b) &-(\bar a\dot a+\bar b\dot b)&\st(a\dot b-\dot ab)\cr \st\cos{\th}(\bar a\dot a+\bar b\dot b)-\dot\th&\st(\dot{\bar a}\bar b-\bar a\dot{\bar b})&\sin^2{\th}(\bar a\dot a+\bar b\dot b)\epm, $$
 $$  P^\perp V(k)= P^\perp (k^{-1}\dot k- (\eta R+\mu R_k)k^{-1}k')=\bpm 0&0&\dot\th\\0&0&0\\-\dot\th&0&0\epm+$${  $$+\st \bpm 0&0&\cos{\th}(\bar a\dot a+\bar b\dot b-3(\eta+\mu))\cr  0&0&(a\dot b-\dot ab-3\mu ab\sin^2{\th})\cr \cos{\th}(\bar a\dot a+\bar b\dot b+3(\eta+\mu)) &(\dot{\bar a}\bar b-\bar a\dot{\bar b}+3\mu \bar a\bar b\sin^2{\th})&0\epm$$}

  Now an element $\chi$ of the $su(3)$ is encoded in a vector
 column $(x_i)$, $i=1,\dots 8$, such that
 $$\chi=\ri\bpm x_3+y&x_1-\ri x_2&x_4-\ri x_5\cr x_1+\ri x_2&-x_3+y&x_6-\ri x_7\cr x_4+\ri x_5&x_6+\ri x_7&-2y\epm.$$
 We have in this basis 
 $$R=\bpm 0&-1&0&0&0&0&0&0\cr 1&0&0&0&0&0&0&0\cr 0&0&0&0&0&0&0&0\cr 0&0&0&0&-1&0&0&0\cr 0&0&0&1&0&0&0&0\cr
   0&0&0&0&0&0&-1&0\cr0&0&0&0&0&1&0&0\cr 0&0&0&0&0&0&0&0\epm,\quad P^\perp=\bpm 0&0&0&0&0&0&0&0\cr 0&0&0&0&0&0&0&0\cr 0&0&0&0&0&0&0&0\cr 0&0&0&1&0&0&0&0\cr 0&0&0&0&1&0&0&0\cr
   0&0&0&0&0&1&0&0\cr0&0&0&0&0&0&1&0\cr 0&0&0&0&0&0&0&0\epm.$$
   Set $$a=a_1+\ri a_2,\quad b=b_1+\ri b_2,\  \  c=\ct,\  s=\st.$$
   Then
   $$\Ad_k=\bpm 2a_1^2+2b_2^2-1&2(a_1a_2-b_1b_2)&2(a_1b_1+a_2b_2)&0&0&0&0&0\cr -2(a_1a_2+b_1b_2)&2a_1^2+2b_1^2-1& 2(a_1b_2-a_2b_1)&0&0&0&0&0\cr 2(a_2b_2-a_1b_1)&-2(a_1b_2+a_2b_1)&2a_1^2+2a_2^2-1&0&0&0&0&0\cr 0&0&0&a_1&a_2&-b_1&b_2&0\cr 0&0&0&-a_2&a_1&-b_2&-b_1&0\cr
   0&0&0&b_1&b_2&a_1&-a_2&0\cr0&0&0&-b_2&b_1&a_2&a_1&0\cr 0&0&0&0&0&0&0&1\epm\times$$$$\times\bpm c&0&0&0&0&s&0&0\cr 0&c&0&0&0&0&-s&0\cr 0&0&1-\frac{s^2}{2}&sc&0&0&0&-\frac{3s^2}{2}\cr 0&0&-sc&1-2s^2&0&0&0&-3sc\cr 0&0&0&0&1&0&0&0\cr
   -s&0&0&0&0&c&0&0\cr0&s&0&0&0&0&c&0\cr 0&0&-\frac{s^2}{2}&sc&0&0&0&1-\frac{3s^2}{2}\epm$$
   Set 
   $$  
   \sigma =\Im(ab)=a_1b_2+a_2b_1,\  \tau=-\Re(ab)=-a_1b_1+a_2b_2, \quad \si^2+\tau^2=\vert a\vert^2\vert b\vert^2. $$
   Then 
  {\small $$R_k= \bpm 0&1-2c^2a\bar a&- c (c^2+1)\sigma&-2c^2s\sigma&0&0&2csa\bar a&3cs^2\sigma\cr 2c^2a\bar a-1&0&-c (c^2+1)\tau&-2c^2s\tau&0&2csa\bar a&0&3cs^2\tau\cr c (c^2+1)\sigma &c (c^2+1)\tau &0&0&cs&s(c^2+1)\sigma&-s(c^2+1)\tau&0\cr 2c^2s\sigma&2c^2s\tau&0&0&s^2-c^2&2cs^2\sigma&-2cs^2\tau&0\cr 0&0&-cs&c^2-s^2&0&0&0&-3cs\cr
   0&-2csa\bar a&-s(c^2+1)\sigma&-2cs^2\sigma&0&0&-1+2s^2a\bar a&3s^3\sigma\cr-2csa\bar a&0&s(c^2+1)\tau&2cs^2\tau&0&1-2s^2a\bar a&0&-3s^3\tau\cr -cs^2\sigma&-cs^2\tau&0&0&cs&-s^3\sigma&s^3\tau&0\epm$$}

   \medskip

   Set 
   $$\gamma=\cos{2\th},\quad \xi=\sin{2\th}$$

Then we find
$$1^\perp-P^\perp (\eta R+\mu R_k)^2P^\perp=\bpm \e_1 &0&p\tau&p\sigma\cr 0&\e_2&-q\sigma&q\tau\cr p\tau&-q\sigma&\bt+\delta \tau^2&\delta\tau\sigma\cr p\sigma&q\tau&\delta \tau\sigma&\bt+\delta\sigma^2\epm=$$
$$=\bpm 1&0&0&0\cr 0&1&0&0\cr 0&0&\sigma&\tau\cr 0&0&-\tau &\sigma\epm\bpm \e_1&0&0&p\cr 0&\e_2&-q&0\cr 0&-q&\frac{\bt}{\tau^2+\sigma^2}&0\cr p&0&0&\frac{\bt}{\tau^2+\sigma^2}+\delta\epm\bpm 1&0&0&0\cr 0&1&0&0\cr 0&0&\sigma&-\tau\cr 0&0&\tau &\sigma\epm$$
where
$$\e_1=1+(\eta+\mu\ga)^2+\mu^2\xi^2(\tau^2+\sigma^2),\quad \e_2=1+\eta^2+\mu^2+2\eta\mu\ga, $$
$$   p=\mu\xi s(\eta-\mu(\vert a\vert^2-\vert b\vert^2)), \quad q=\eta\mu \xi s, \quad \beta=1+(\eta+\mu)^2-4\eta\mu s^2\vert a\vert^2,\quad \delta=-4\mu^2s^2. $$
For the inverse matrix we find 
$$(1^\perp-P^\perp (\eta R+\mu R_k)^2P^\perp)^{-1}=$$$$=\bpm 1&0&0&0\cr 0&1&0&0\cr 0&0&\frac{\sigma}{\sigma^2+\tau^2}&\frac{\tau}{\sigma^2+\tau^2}\cr 0&0&-\frac{\tau}{\sigma^2+\tau^2}&\frac{\sigma}{\sigma^2+\tau^2}\epm\bpm \Delta_1(\frac{\bt}{\sigma^2+\tau^2}+\delta) &0&0&\Delta_1 p \cr 0&\frac{\Delta_2\bt}{\sigma^2+\tau^2}&\Delta_2q&0\cr 0&\Delta_2q&\Delta_2\e_2&0\cr \Delta_1 p &0&0&\Delta_1\e_1  \epm\bpm 1&0&0&0\cr 0&1&0&0\cr 0&0&\frac{\sigma}{\sigma^2+\tau^2}&-\frac{\tau}{\sigma^2+\tau^2}\cr 0&0&\frac{\tau}{\sigma^2+\tau^2}&\frac{\sigma}{\sigma^2+\tau^2}\epm$$
where
$$\Delta_1=\frac{\sigma^2+\tau^2}{\e_1\bt+(\e_1\delta-p^2)(\sigma^2+\tau^2) }, \quad \Delta_2=\frac{\sigma^2+\tau^2}{\e_2\bt -q^2(\sigma^2+\tau^2) }.$$
We have
$$P^\perp V(k)=\bpm V_4(k)\cr V_5(k)\cr V_6(k)\cr V_7(k)\epm=\bpm sc\Im(\bar a\dot a+\bar b\dot b)
\cr \dot\th+sc\Re(\bar a\dot a+\bar b\dot b)-3(\eta+\mu)sc\cr
 s\Im(a\dot b-\dot ab)-3\mu \sigma s^3\cr s\Re(a\dot b-\dot ab)+3\mu \tau s^3\epm.$$
 This gives
\be -\jp\tr (P^\perp V(k)(1^\perp-P^\perp (\eta R+\mu R_k)^2P^\perp)^{-1}P^\perp V(k))=$$$$= \bpm W_4(k)&W_5(k)&W_6(k)&W_7(k)\epm\bpm \Delta_1(\frac{\bt}{\sigma^2+\tau^2}+\delta) &0&0&\Delta_1 p \cr 0&\frac{\Delta_2\bt}{\sigma^2+\tau^2}&\Delta_2q&0\cr 0&\Delta_2q&\Delta_2\e_2&0\cr \Delta_1 p &0&0&\Delta_1\e_1  \epm \bpm W_4(k)\cr W_5(k)\cr W_6(k)\cr W_7(k)\epm,\label{pvv}\ee
 where
 \be \bpm W_4(k)\cr W_5(k)\cr W_6(k)\cr W_7(k)\epm=\bpm 1&0&0&0\cr 0&1&0&0\cr 0&0&\frac{\sigma}{\sigma^2+\tau^2}&-\frac{\tau}{\sigma^2+\tau^2}\cr 0&0&\frac{\tau}{\sigma^2+\tau^2}&\frac{\sigma}{\sigma^2+\tau^2}\epm \bpm V_4(k)\cr V_5(k)\cr V_6(k)\cr V_7(k)\epm=\bpm sc\Im(\bar a\dot a+\bar b\dot b)\cr\dot\th-3sc(\eta+\mu)\cr s\Re\left(\frac{\dot  b }{b}-\frac{\dot{ a}}{a}\right)+3\mu s^3\cr -s\Im\left(\frac{\dot  b }{b}-\frac{\dot{ a}}{a}\right)\epm.\label{wv}\ee
 Inserting \eqref{pvv} into the general bi-YB action \eqref{sor} gives the $\cp^2$ doubly deformed action \eqref{3so}
 $$    S_{\eta,\mu}(k)= \int dt\   \Delta_2\left(\frac{\bt}{\vert a\vert^2\vert b\vert ^2} W_5^2+\e_2 W_6^2+2qW_5W_6\right)+$$
$$\int dt  \Delta_1\left(\left(\frac{\bt}{\vert a\vert^2\vert b\vert ^2}+\delta\right)W_4^2+\e_1 W_7^2+2pW_4W_7\right) -9\int dt \sin^2{\th}    $$ presented  in Section 6.3.

\medskip 

 For $\mu=\eta=0$, the  equations \eqref{pvv}, \eqref{wv} become
 \be -\jp\tr (P^\perp V(k)P^\perp V(k))=W_4^2+W_5^2+\vert a\vert^2\vert b\vert^2(W_6^2+W_7^2),\label{nn}\ee
  $$\bpm W_4(k)\cr W_5(k)\cr W_6(k)\cr W_7(k)\epm=\bpm 1&0&0&0\cr 0&1&0&0\cr 0&0&\frac{\sigma}{\sigma^2+\tau^2}&-\frac{\tau}{\sigma^2+\tau^2}\cr 0&0&\frac{\tau}{\sigma^2+\tau^2}&\frac{\sigma}{\sigma^2+\tau^2}\epm \bpm V_4(k)\cr V_5(k)\cr V_6(k)\cr V_7(k)\epm=\bpm sc\Im(\bar a\dot a+\bar b\dot b)\cr\dot\th \cr s\Re\left(\frac{\dot  b }{b}-\frac{\dot{ a}}{a}\right)\cr -s\Im\left(\frac{\dot  b }{b}-\frac{\dot{ a}}{a}\right)\epm,$$
   which can be checked to correspond to the non-deformed Fubini-Study metric on $\cp^2$.


 \end{document}